\documentstyle[12pt]{article}
\begin{titlepage}   
\title{ 
{\vspace{-8.5ex}
\normalsize\bf
\begin{flushright}
IHEP 99--12\\
hep-ph/9903344\\
\end{flushright}
\vspace{10ex}
}
{\bf Vector-like family extension }\\
{\bf of the standard model and}\\
{\bf the light quark masses}\\
{\bf and mixings}\\[3ex]
}
                                                                     \author{ 
{Yury F.~Pirogov}$^1$ \and Oleg V.~Zenin\\[1ex]
{\it Institute for High Energy Physics,}\\
{\it Protvino, RU-142284 Moscow Region, Russia}\\[0.5ex]
{\it  Moscow  Institute  of  Physics  and  Technology,}\\
{\it  Dolgoprudny,  Moscow  Region, Russia }}
\date{}

\begin{document}
\maketitle
\thispagestyle{empty}
\addtocounter{page}{1}
\addtocounter{footnote}{1}

\abstract{
\noindent
The standard model extended with  the pairs  of the vector-like
families is studied. 
The model independent analysis for an arbitrary case and  an
explicit realization for the case with one pair of the heavy
vector-like
families are considered. The mixing matrices 
of the light quarks for the  left- and
right-chiral charged currents, as well as those for the flavour
changing neutral currents, both the $Z$ and Higgs mediated,  are
found. 
}
{
\vspace{8ex}
\begin{flushleft}
$^1$ E-mail address: pirogov@mx.ihep.su   
\end{flushleft}
}
\end{titlepage}

\paragraph*{1. Introduction }

Are there extra families in the standard model (SM) or not 
this is the question.
A recent two-loop renormalization group analysis~\cite{pir} of the SM
shows that subject to
the precision experiment restriction on the Higgs
mass, $M_H\le 200$~GeV at 95\%~C.L.~\cite{tournefier}, the forth
chiral family, if alone, is excluded.\footnote{More conservative
restrictions 
$m_H\le 262$ GeV or $M_H\le 300$ GeV at $95\%$ C.L., respectively, 
from the first and second papers of Ref.~\cite{higgs} though render
this conclusion somewhat less reliable, nevertheless do not
invalidate it.} 
In fact, it does not depend on whether
this extra family has the normal chiral structure or the mirror one. 
But as it  is noted 
in the Ref.~\cite{pir},
a pair of the opposite chirality  families with the 
relatively low Yukawa couplings evades the SM self-consistency
restrictions and could still exist. In order to
conform with observations these extra families, which otherwise can
be considered as the vectorial ones,  should get large
direct masses and to drop out of the light particle spectrum of the
SM in the decoupling limit. 
Nevertheless, 
at the moderate masses, say of order 1 TeV or so,
such families  could lead to observable
corrections to the SM interactions through mixing with  the light
fermions. 

The various vector-like fermions are generic  in many extensions of
the SM like the superstring and grand unified theories, composite
models, etc. Many issues concerning those fermions, both the
electroweak doublets and singlets, the latter ones of the up and down
types, 
were considered in the literature~\cite{vlf},~\cite{lavoura}. 
On the other hand there are numerous studies of the $n>3$ chiral
family extensions of the SM~\cite{n_families}, \cite{santa}. 
Some topics concerning the SM extensions with the vector-like
families are studied in Ref.~\cite{fuji}.
But the problem of the SM quark masses and mixings in the presence of
the extra vector-like families have not yet found its
full model independent consideration, and it is
studied in the current paper.  We present both the model independent
analysis for the general case and an explicit realization for the
case with a pair of the heavy vector-like families.

\paragraph*{2. Arbitrary number of the vector-like families }

The most general content of the SM families  consisting of the
$\mathrm SU(2)_W \times U(1)_Y$  doublets and singlets  
in the chiral notations is $nQ_L + mQ'_R$, where
$Q_L = ({\hat q}_L, {\hat u}^c_L, {\hat d}^c_L)$ and
$Q_R = ({\hat q}_R', {\hat u}'^c_R, {\hat d}'^c_R)$.
The symbols with the hat sign designate quarks in the
symmetry/electroweak
basis where, by definition, the SM symmetry structure is well stated.
Here $n\ge 3$ is the number of chiral families, similar in
their chiral and quantum number
structure to three ordinary families of the minimal SM. 
$m\ge 0$ means the number of the mirror conjugate families 
with the normal quantum
numbers, or in other terms, the charge conjugate families with the
normal chiral structure.  
In the more traditional left-right notations, one should substitute:
$Q_L\to ({\hat q}_L, {\hat u}_R, {\hat d}_R)$ and
$Q_R\to ({\hat q}'_R, {\hat u}'_L, {\hat d}'_L)$.

In general, quarks gain masses from two different physical
mechanisms: that of the SM Yukawa interactions and that of a New
Physics resulting in the SM invariant direct mass terms.
Being chirally unprotected the latter ones should naturally 
be characterized by a high mass scale $M$, $M\gg v$, with $v$
being the SM Higgs vacuum expectation value.
In the symmetry basis the kinetic, Yukawa and direct mass
Lagrangian has the following most general form:
\def\D{D\hspace{-0.28cm}/\hspace{0.08cm}}
\begin{eqnarray}
\label{eq:lagrangian}
{\cal L}&=&
~~~i \overline{{\hat q}_L}\D {\hat q}_L + i \overline{{\hat u}_R}\D
{\hat u}_R + 
i \overline{{\hat d}_R}\D {\hat d}_R\nonumber\\
&&+\, i\overline{{\hat q}'_R}\D {\hat q}'_R + i \overline{{\hat
u}'_L}\D {\hat u}'_L + 
i \overline{{\hat d}'_L}\D {\hat d}'_L\nonumber\\
&&-\Big{(}\overline{{\hat q}_L} Y^u {\hat u}_R \phi^c +
\overline{{\hat q}_L} Y^d {\hat d}_R\phi
+\overline{{\hat u}'_L} {Y^u}' {\hat q}'_R {\phi^c}^{\dagger} 
+ \overline{{\hat d}'_L} {Y^d}' {\hat q}'_R \phi^{\dagger} 
+ \mbox{h.c.}\Big{)}
\nonumber\\
&& - \Big{(}
 \overline{{\hat q}_L} M {\hat q}'_R + 
\overline{{\hat u}'_L}{M^u}' {\hat u}_R
+ \overline{{\hat d}'_L} {M^d}' {\hat d}_R + \mbox{h.c.}\Big{)}~,
\end{eqnarray}
\noindent
where $\D\equiv \gamma^\mu D_\mu$ is the SM covariant derivative and 
$\phi$ is the Higgs doublet and $\phi^c$ is the garged conjugate one.
In Eq.~(\ref{eq:lagrangian}),
$Y$ and  $Y'$ are, respectively,
the square $n\times n$ and $m\times m$ Yukawa matrices; 
$M$ and  $M'$ are, respectively, the rectangular $n\times m$ and
$m\times n$ direct mass matrices. 

We generalize the parameter counting for the
chiral families of Ref.~\cite{santa} to the case with the extra 
vector-like families (VLF's).
It goes as is shown in Table 1. Here $G$ is the global symmetry of
the kinetic part of the Lagrangian~(\ref{eq:lagrangian}). It is
broken explicitly by the mass terms, only the residual symmetry $H =
U(1)$ of the baryon number  being left in
the  general case we consider.
Hence, the transformations  of
$G/H$ can be used to absorb the spurious parameters in
Eq.~(\ref{eq:lagrangian})
leaving only the physical set ${\cal M}_{phys}$ of them. 
Of the physical moduli, the $2(n + m)$ ones are the physical masses,
the rest being mixing angles. The last
two lines in Table 1 present the physical parameters for the
minimal SM and for its extension with a pair of the normal and mirror 
families. This case will be considered in detail further on.

\begin{table}[htbp]
\paragraph{Table 1}Parameter counting in the symmetry/electroweak
basis.
\vspace{1ex}

\begin{tabular}{|c|c|c|}
\hline 
Couplings&Moduli&Phases\\
and symmetries&&\\
\hline
$Y^u, Y^d, {Y^u}', {Y^d}',$&$2~(n^2 + m^2)$&$2~(n^2 + m^2)$\\ 
$M, {M^u}', {M^d}'$&$+ 3~n m$&$+ 3~n m$\\
\hline
$G = U(n)^3\times U(m)^3$&$-\frac{3}{2} [n(n-1) + m(m-1)]$&
$-\frac{3}{2} [n(n+1) + m(m+1)]$\\  
\hline
$H = U(1)$&$0$&$1$\\
\hline
${\cal M}_{phys}(n, m)$&$\frac{1}{2} (n+m)(n+m-1) $&
$\frac{1}{2} (n+m-2)(n+m-1)$\\
&$ + 2~n m +~2~(n + m)$&$ +2~n m$\\
\hline
${\cal M}_{phys}^{SM}(3, 0)$&$6 +3 =  9  $&$1$\\
\hline
${\cal M}_{phys}(4, 1)$&$10 + 18 = 28$&$14$\\
\hline
\end{tabular}
\end{table}

Let us now redefine collectively quarks 
in the symmetry basis as
\def\k{\kappa}
${\hat \k}_\chi = {\hat u}_\chi$, ${\hat d}_\chi$
and these in  the mass basis, i.e.\ the quark eigenstates with
${\cal M}_{phys}$ being diagonal, as $\k_\chi = u_\chi$, $d_\chi$
($\chi =
L$, $R$). The bases are related by the unitary $(n+m)\times(n+m)$
transformations

\begin{equation}
{\hat\k}_\chi{}_{A} = 
({U^\k_\chi})^{ F}_{ A}
\,{\k_\chi}_{F}~,
\end{equation}
with the ensuing bi-unitary mass diagonalization
\begin{equation}\label{eq:b_tr}
{U^{\k}_L}^\dagger {\cal M}^{k} U^{\k}_R = 
{\cal M}^{\k}_{diag} = \mbox{diag\,}(\overline{ m}^{\k}{}_f , 
\overline{ M}^{\k}{}_4, \dots ,\overline{ M}^{\k}{}_{n+m})~. 
\end{equation}
In equations above, the indices  $A = A_L,A_R$; $A_L = 1,\dots, n$;
$A_R = n+1,\dots, n+m$ are those in the symmetry basis, and 
$F = f,4, \dots ,n+m$; $f = 1,2,3$ are indices in the mass basis.
It is assumed that $\overline{m}^\k{}_f \ll \overline{M}^\k{}_4,
\dots , \overline{ M}^\k{}_{n+m}$.

The matrices $U^\k_\chi$ satisfy the unitarity relations 
\begin{equation}
\label{6}
{U^{\k}_\chi}\,U^{\k}_\chi{}^{\dagger} =I
\end{equation} 
and
\begin{equation}\label{5}
{U^{\k}_\chi}^{\dagger} I_L {U^{\k}_\chi} 
+ {U^{\k}_\chi}^{\dagger}I_R  {U^{\k}_\chi} = I~,
\end{equation}
were $I_L$, $I_R$ are the projectors onto the normal and mirror
subspaces in the symmetry basis:
\begin{eqnarray}
I_L&=&\mbox{diag}\,
(\,\underbrace{1,\dots,1}_{n}\,;\underbrace{0,\dots,0}_{m}\,)~,
\nonumber\\
I_R&=&\mbox{diag}\,
(\,\underbrace{0,\dots,0}_{n}\,;\underbrace{1,\dots,1}_{m}\,)
\end{eqnarray}
with $I_L+I_R=I$ and $I_\chi^2=I_\chi$.
Let us also introduce their transforming to the mass basis
\begin{equation}
\label{eq:projector}
X^\k_\chi = U^\k_\chi{}^{\dagger} I_\chi U^\k_\chi~.
\end{equation}
($\k = u$, $d$ and  $\chi = L$, $R$). Clearly, $X^\k_\chi$ are
Hermitian
and satisfy the projector condition: 
$X^\k_\chi{}^2 = X^\k_\chi{}$ (but
note that $X_L^\k + X_R^\k\neq I$ in the notations adopted). 

Now, the charged current Lagrangian is
\begin{equation}\label{eq:L_W}
- {\cal L}_W = \frac{g}{\surd\overline{2}} W^+_\mu 
\sum_\chi \overline{u_\chi} \gamma^\mu V_\chi d_\chi + \mbox{h.c.}
\end{equation}
and the neutral current one is
\begin{equation}\label{eq:L_Z}
- {\cal L}_Z = \frac{g}{c} Z_\mu
\sum_{\k,\chi} \overline{\k_\chi} \gamma^\mu N^{\k}_\chi\, \k_\chi~,
\end{equation}
where $c\equiv\cos\theta_W$, with $\theta_W$ being the 
Weinberg mixing angle. 
The corresponding quark mixing matrices  for the charged
currents are:
\begin{equation}\label{eq:V_chi}
{V_\chi} =
{U^{u}_\chi}^{\dagger}I_\chi U^{d}_\chi~,
\end{equation}
and for the neutral currents with the operator $T_3-s^2 Q$:
\begin{equation}\label{eq:N_chi}
N^\k_\chi=T^\k_3 X^\k_\chi-s^2 Q^\k_\chi~.
\end{equation}
Here $T_3^\kappa$ is the 3rd component of the electroweak isospin for 
$\kappa = u, d$ and $Q_{L,R}^\kappa\equiv Q^\kappa I$, with
$Q^\kappa$
being the corresponding electric charge; $s\equiv \sin \theta_W$.

The charged current mixing matrices 
$V_L$ and $V_R$ play the role of the generalized
CKM matrices. But contrary to the minimal SM case, they as well as
the neutral current mixing matrices $N^{\k}_\chi$ are non-unitary.
Namely, one gets by the unitarity relations~(\ref{6})
\begin{eqnarray}\label{14}
V_\chi V_\chi^\dagger &=&  X^u_\chi ~,\nonumber\\
V_\chi^\dagger V_\chi &=& X^d_\chi~, 
\end{eqnarray}
where $X_\chi^\k$ ($X_\chi^\k\neq I$ in general) are given by
Eq.~(\ref{eq:projector}). 

It is seen that the neutral current matrices  $N^\k_\chi$ are not
independent of the charged current ones $V_\chi$. In fact, one can
convince oneself  that $V_\chi$ and the diagonal mass matrices
${\cal M}^{\k}_{diag}$ suffice to parametrize all the fermion
interactions in a general class of the SM extensions by means 
of the arbitrary numbers 
of the vector-like isodoublets and
isosinglets~\cite{lavoura}.
Indeed, in the case at hand using the unitarity relations (\ref{5})
one gets for the Yukawa Lagrangian in the unitary gauge
\begin{eqnarray}\label{L_Y}
- {\cal L}_Y &=&
\frac{H}{v} \sum_\k \overline{\k_L} \Big{(} X^\k_L {\cal M}^\k_{diag}
-2 X^\k_L {\cal M}^\k_{diag} X^\k_R  
+ {\cal M}^{\k}_{diag} X^\k_R \Big{)} \k_R 
\nonumber\\
&& + \sum_\k \overline{\k_L} {\cal M}^{\k}_{diag} \k_R +
\mbox{h.c.}~,
\end{eqnarray}
$H$ being the physical Higgs boson. 
It follows from the above expression
and Eqs.~(\ref{eq:L_Z}), (\ref{eq:N_chi})
that all the flavour changing neutral currents are induced entirely 
by the lack of unitarity of the charged current mixing matrices
$V_\chi$. In the case with only the normal
families  ($X^\k_L=I$, $X^\k_R=0$) the usual SM  expressions for
${\cal L}_W$, ${\cal L}_Z$ and ${\cal L}_Y$ are recovered, the two
latter ones being flavour conserving.

We propose the following prescription for the model independent
paramet\-ri\-za\-tion of the $V_\chi$. 
The problem is that they are non-unitary and thus are difficult 
to parametrize directly. So, the idea is to express them in 
terms of a set of the auxiliary unitary matrices.
First of all, note that in absence of any restrictions on the
Lagrangian
the unitary matrices $U^\kappa_\chi$ in Eq.~(2) would be arbitrary.
Now, an arbitrary
$(n + m)\times (n + m)$ unitary matrix $U$ can always be uniquely 
decomposed as
$U = U{\vert}_{n\times n} ~U{\vert}_{m\times m} ~U{\vert}_{n\times
m}$. 
Here $U{\vert}_{n\times n}$ is a unitary matrix in the $n\times n$ 
subspace. It is built of the $n^2$ generators. Similarly, 
$U{\vert}_{m\times m}$ is the restriction of $U$ onto the $m\times m$
subspace, and it is built of the $m^2$ generators. And finally,
$U{\vert}_{n\times m}$ means a unitary $(n + m)\times (n + m)$
matrix built of the $2nm$ generators which mix the two subspaces.

Now, by means of the symmetry basis transformations $G$ of the Table
1 
one can always put, without loss of generality, the matrices
$U^\kappa_\chi$ to the form
\begin{eqnarray}\label{eq:U_repr}
U^u_L &=& U^u_L{\vert}_{n\times m}~,
\nonumber\\
U^u_R &=& U^u_R{\vert}_{n\times m}~,
\nonumber\\
U^d_L &=& U^d_L{\vert}_{n\times n} ~U^d_L{\vert}_{n\times m}~,
\nonumber\\
U^d_R &=& U^d_R{\vert}_{m\times m} ~U^d_R{\vert}_{n\times m}~.
\end{eqnarray}
This representation includes six auxiliary unitary matrices.
Clearly, they depend on the $[n(n-1)/2 + m(m-1)/2 + 4mn]$ moduli and
$[n(n+1)/2 + m(m+1)/2 + 4mn]$ phases, and these numbers are
redundant.
But the $nm$ moduli and the same number of phases can be eliminated 
through the $n\times m$ matrix constraint
\begin{equation}\label{eq:constraint}
I_L U^u_L {\cal M}^u_{diag} {U^u_R}^{\dagger} I_R =
I_L U^d_L {\cal M}^d_{diag} {U^d_R}^{\dagger} I_R~.
\end{equation}
The latter follows from the equality of the direct mass matrices $M$
in Eq.~(1) for the up and down quarks, 
and it includes additionally the $2(n+m)$ independent
moduli which enter ${\cal M}^u_{diag}$ and ${\cal M}^d_{diag}$.
By means of the Eq.~(\ref{eq:constraint}) one can express, 
e.g., one of the 
$U^\kappa_\chi{\vert}_{n\times m}$ in terms of all other matrices.
And finally, the $2(n+m)-1$ phases can be removed via the residual
phase
redefinition for the quarks in the mass basis. Putting all together,
one can easily verify that the total number of the independent
parameters is precisely as expected from the Table 1.

Having parametrized the auxiliary unitary matrices, 
one gets for the $V_\chi$
\begin{eqnarray}\label{eq:V_repr}
V_L &=& {U^u_L}^\dagger{\vert}_{n\times m}~ I_L~
        {U^d_L}{\vert}_{n\times n} ~{U^d_L}{\vert}_{n\times m}~,
\nonumber\\
V_R &=& {U^u_R}^\dagger{\vert}_{n\times m}~ I_R~
        {U^d_R}{\vert}_{m\times m} ~{U^d_R}{\vert}_{n\times m}
\end{eqnarray}
and for the $X^\kappa_\chi$
\begin{equation}\label{eq:X_repr}
X^\kappa_\chi = {U^\kappa_\chi}^\dagger{\vert}_{n\times m}~ I_\chi~ 
                {U^\kappa_\chi}{\vert}_{n\times m}~.
\end{equation}
When eliminating the $2(n+m)-1$ redundant phases one can always
take such a choice as to render the diagonal and above-the-diagonal
elements of the $V_L$ (or $V_R$) to be real and positive.

This gives the principal solution to the problem. When there are only 
the normal families ($m = 0$) the usual parametrization in terms of
just one unitary matrix $U^d_L{\vert}_{n\times n}$ is readily
recovered.
For the case with a pair of VLF's ($n=4$, $m=1$) we got also the
explicit expressions
of all the relevant quantities in terms of a minimal common set of
the independent arguments parametrizing the mass matrices (see
below).

\paragraph*{3. A pair of the heavy vector-like families }

The mass/flavour basis quantities, ${\cal M}^{u,d}_{diag}$ and
$V_{L,R}$, are
phenomenological by their very nature.
They reflect an obscure mixture of contributions of the quite
different physical origin. In particular, they shed no light on
the mixing magnitudes. On the contrary, the parameters in the
symmetry basis,
i.e.\ Yukawa couplings and the direct mass terms $M$ and
${M^u}'$, ${M^d}'$
have the straightforward theoretical meaning. So, we express the
former ones in terms of the latter ones. This permits us to expand
upon the idea of the relative magnitude of the various mixing
elements in terms of the small quantity $v/M$.

The asymptotic freedom requirement for the $\mathrm SU(2)_W$
electroweak interactions results in the restriction that the total
number of the electroweak doublets 
should not exceed 21, and thus the total number of 
the families is $(n+m)\le 5$. 
Hence the maximum number of the extra VLF's allowed by the asymptotic
freedom is two, the case we stick to in what follows.

Using here the global symmetries $G$ of the Table~1 one can bring,
without loss of generality,  the
quark mass matrices in the symmetry basis to the following canonical
form 
\begin{equation}\label{20}
{\cal M}^\k = \left(
\begin{array}{ccc}
\vspace{0.5ex}
{m^\k}^g_f&{\mu^\k}'_f&0\\
\vspace{0.5ex}
{\mu^\k}^g&m^\k{}_4&M\\
0&{M^\k}'&m^\k{}_5
\end{array}
\right)~,
\end{equation}
where $M$, ${M^\k}'$ are the real scalars and $\mu^\k{}^f$,
${\mu^\k}'_f$,
$m^\k{}_4$, $m^\k{}_5$ are in general complex. 
Here the lower case characters generically mean the masses of the
Yukawa
origin ($\sim Yv$).
Let us remind that $M$ in Eq.~(\ref{20}) is common for both
${\cal M}^u$ and ${\cal M}^d$.
The three-dimensional matrices $m^\k$ are Hermitian and positive
definite,  and one of them, e.g. $m^u$, can always be  chosen
diagonal. Under such a choice one can  simplify further:
\begin{equation}\label{21}
{\cal M}^\k_0 = {U^\k_0}^\dagger {\cal M}^\k U^\k_0, 
\end{equation}
where
\begin{equation}\label{eq:M_0_kappa}
{\cal M}_0^{\k} = \left(\begin{array}{ccccc}
\vspace{0.5ex}
m^\k{}_1&0&0&{\mu^\k}'_1&0\\
\vspace{0.8ex}
0&m^\k{}_2&0&{\mu^\k}'_2&0\\
\vspace{0.5ex}
0&0&m^\k{}_3&{\mu^\k}'_3&0\\
\vspace{0.5ex}
{\mu^\k}^1&{\mu^\k}^2&{\mu^\k}^3&m^\k{}_4&M\\
\vspace{0.5ex}
0&0&0&{M^\k}'&m^\k{}_5     
\end{array}\right)
\end{equation}
\noindent
with a redefinition of $\mu^\k{}^f$ and ${\mu^\k}'_f$, and with the
diagonal elements $m^\k{}_f$ being real and positive. 
The corresponding unitary $U^\k_0$ are given by 
\begin{eqnarray}\label{22a}
U^{u}_0 &=& I~,\nonumber\\
U^d_{0 } &=& \left({
\begin{array}{cc}
V_{C}&0\\
0&I_2
\end{array}
}\right)~,
\end{eqnarray}
$V_C$ being the $3\times 3$ CKM matrix  and $I_2$ the $2\times 2$
identity
matrix.  The mass matrices of
Eq.~(\ref{eq:M_0_kappa}) possess the residual symmetry $U(1)^6$ which
is reduced
to  $U(1)^5$  by the baryon number conservation.
So, one can use phase redefinitions for two of the light
$d$ quarks 
which leave just one complex phase in $V_C$ in accordance with the
decoupling limit requirement. 

It is seen from Eqs.~(\ref{eq:M_0_kappa}) and (\ref{22a}) that in
this
parametrization 
the total number of the physical moduli is $10 + 15 + 3 = 28$ 
as it should be according to the Table 1.
What concerns the phases, their number is in general $16 + 1 = 17$,
i.e.\ three of them are spurious
and can be removed. 
E.g., by means of the residual phase redefinition for the  three
light $u$ quarks one can make $\mu^u{}^f$ or ${\mu^u}'_f$ to be real,
or put some other three relations on their phases.
This exhausts the freedom of the phase redefinitions, leaving only
the physical parameters. 

Solving the characteristic equations 
$\det\,({\cal M}^\k_0 {{\cal M}^\k_0}^\dagger -
{\overline{m}^\kappa}^2 I) = 0$
one gets for the light quark physical masses in the first order 
(i.e. up to the relative
corrections ${\cal O}(v^2/M^2)$ to the leading order):
\begin{equation}\label{eq:lambda}
\overline{ m}_f^2 = m_f^2\bigg{(}1
 - \Big(\frac{\vert\mu^f\vert^2}{M^2} +
 \frac{\vert\mu'_f\vert^2}{M^{'2}}\Big)\bigg{)}
 + \frac{m_f}{M M'} (m_5\mu^f\mu'_f + \mbox{h.c.})
\end{equation}
with the superscripts $\k = u$, $d$ being suppressed.
Here it is supposed that $M\sim M'$ but $M\neq M'$ in general.
It is seen that corrections to $m_f^2$ are proportional to $m_f$ 
themselves, i.e. the light quarks are chirally protected.
This property drastically reduces the otherwise dangerous corrections
to the masses of the lightest $u$ and $d$ quarks at the moderate $M$.
On the other hand, it means that the masses of the lightest quarks
can not
entirely be induced by an admixture of the vector-like families:
if $m_f = 0$ then $\overline{m}_f = 0$, too.

Once the physical masses are known, one can obtain the matrices
$U^\k_L$ and 
$U^\k_R$ of the bi-unitary transformation~(\ref{eq:b_tr}).
With account for Eq.~(\ref{eq:V_chi})
one gets hereof for the light quark mixing matrix $V_L$ 
\def\u{{p'}^u{}}
\def\d{{p'}^d{}}
\begin{equation}\label{eq:DeltaV_L}
{V_L}{}^g_f = {V_C}{}^g_f \Big{(} 1
- \frac{1}{2 M^2} ({n^u}{}^f_f + {n^d}{}^g_g) \Big{)} 
- \frac{1}{M^2} \sum ({{p^u}^f_h}^* {V_C}^g_h + {V_C}_f^h {p^d}^g_h)
\end{equation}
and similarly for $V_R$
\def\u{{p^u}'{}}
\def\d{{p^d}'{}}
\begin{equation}\label{eq:V_R}
{V_R}{}^g_f = \frac{1}{{M^u}' {M^d}'} {{\u}^f_5}^* {\d}^g_5~,
\end{equation}
where
\begin{eqnarray}\label{eq:p}
p^f_g &=&                          
\frac{\mu^f (m_f^2-\vert m_5\vert^2)(m_f {\mu^f}{}^*\mu'_g -
   m_g {\mu^g}^*\mu'_f)
         + k_f (m_f\mu'_g - \frac{m_g}{m_f}\frac{M'}{M} {\mu^g}^*
         m_5^*)}
       {(m_g^2-m_f^2)(m_f\mu'_f - \frac{M'}{M} m_5^*
       {\mu^f}{}^*)}~,\nonumber\\ 
\vspace{0.5ex}\nonumber\\		
p^f_5 &=&
\frac{\frac{M'}{M} (k_f+m_f^2\vert\mu^f\vert^2) - m_f m_5\mu^f\mu'_f}
     {m_f(m_f\mu'_f - \frac{M'}{M} m_5^* {\mu^f}{}^*)}~,\nonumber\\
\vspace{0.5ex}\nonumber\\
n^f_f &=&
\bigg{\vert} 
\frac{\frac{M'}{M} (k_f+m_f^2\vert\mu^f \vert^2) 
       - m_f m_5\mu^f\mu'_f}
     {m_f(m_f\mu'_f - \frac{M'}{M} m_5^* {\mu^f}{}^*)}
\bigg{\vert}^2
\end{eqnarray}
with $k_f = M^2 (\overline{m}_f^2 - m_f^2)$.
The $p'$, $n'$ are obtained from $p$, $n$, respectively, by 
substituting $\mu^f\leftrightarrow {\mu'_f}^*$, 
$m_4\leftrightarrow m_4^*$, $m_5\leftrightarrow m_5^*$,
$M\leftrightarrow M'$. 
All these auxiliary parameters are in general of order ${\cal
O}(M^0)$. 

The charged current Lagrangian ${\cal L}_W$ is given by
Eq.~(\ref{eq:L_W}).
The $Z$ mediated neutral current Lagrangian ${\cal L}_Z$ is as given
by Eqs.~(\ref{eq:L_Z}), (\ref{eq:N_chi}) with 
\def\p{p}
\begin{equation}\label{eq:X_L}
{X_L}{}^g_f = \delta^g_f - \frac{1}{M^2} {\p^f_5}^* \p^g_5
\end{equation}
and
\def\pk{{p'}{}}
\begin{equation}\label{eq:X_R}
{X_R}{}^g_f = \frac{1}{{M'}^2} {{p'}{}^f_5}^* {p'}{}^g_5~.
\end{equation}

The neutral scalar current Lagrangian takes the general form
\begin{equation}\label{29}
- {\cal L}_H = \frac{H}{v} \sum_\k 
\overline{\k_L}\, {U^\k_L}^\dagger ({\cal M}^\k 
- {\cal M}^\k_{dir}) U^\k_R \, \k_R + \mbox{h.c.}
\end{equation}
with the direct mass matrices
\begin{equation}\label{30}
{\cal M}^\k_{dir} = \left({
\begin{array}{ccc}
 O_3&0&0\\
0&0&M\\
0&{M^\k}'&0
\end{array}
}\right)~,
\end{equation}
where $O_3$ is the $3\times 3$ zero matrix. As a consequence
of the
substraction of the direct mass terms, the total mass and Yukawa
matrices are not diagonalizable simultaneously in the same basis.
In the mass basis,  the Higgs interaction Lagrangian is non-diagonal 
\begin{eqnarray}\label{eq:Higgs_Lagr}
- {\cal L}_H &=& \frac{H}{v} \sum_\k {\overline{\k_L}}~ {\cal H}^\k
\k_R + \mbox{h.c.}~,
\end{eqnarray}
with the light quark mixing matrix (indices $\k = u$, $d$ being
omitted) 
\begin{equation}\label{eq:H}
{\cal H}^g_f = \overline{m}_f\delta^g_f  
- \frac{1}{MM'} \bigg{(} {p^f_4}^* {p'}^g_5 + {p^f_5}^*
{p'}^g_4\bigg{)}~,
\end{equation}
where
\begin{equation}\label{eq:p_f_4}
p^f_4 =
- k_f ~\frac{\Big(k_f+\vert\mu^f\vert^2 (m_f^2-\vert m_5\vert^2)
             \Big)\Big(\frac{M'}{M} m_5^*
             +\frac{1}{k_f} m_f\mu^f\mu'_f (m_f^2-\vert
             m_5\vert^2)\Big)} 
       {m_f(m_f\mu'_f - \frac{M'}{M} m_5^*\mu_f^*)}
\end{equation}
with ${p'}^f_4$ being obtained from it by the usual substitutions.

One should stress  that for the light quarks all the off-diagonal 
components of the  Lagrangian  ${\cal L}_W$ (beyond that of the
minimal SM), as well as those of the
${\cal L}_Z$ and ${\cal L}_H$ are suppressed by the ratio $v^2/M^2$,
and it does not depend on the details of the mass matrices.

\paragraph*{4. Conclusions}

We have shown that the mere addition of a pair of the VLF's
drastically changes
all the characteristic features of the minimal SM. First of all, the
generalized CKM matrix for the  left-handed charged currents ceases
to be unitary. Moreover, this non-unitarity takes place in the whole 
flavour space but not only in the light quark sector which would
occur for adding  only 
the normal families.
Further, there appear the right-handed charged currents,
the flavour changing neutral currents, both the vector and  
scalar ones, all with the non-unitary mixing matrices and with a
number of $CP$ violating phases.

Due to decoupling under the large direct mass terms $M$, the
extended
SM definitely does not contradict to experiment in the 
limit $M\gg v$. But at the moderate  $M>v$, the addition of a pair of
the VLF's would make the model
phenomenology, especially that of the flavour and
$CP$  violation, extremely reach. It is to be
seen  what is the real experimentally allowed region  in
the parameter space for the VLF's and what are the possibilities to
observe their effects in the future experiments. We hope that our
paper will stimulate further study in this direction.


\begin{thebibliography}{**}

\bibitem{pir}
Yu.F.\  Pirogov and O.V.\ Zenin, IHEP preprint IHEP 98--50 (1998), 
to~be publ.\ in Eur.\ Phys.\ J.\ C, 1999, 
hep-ph/9808396; Report presented at the Int.\ Seminar
``Quarks '98'', Suzdal, May 17-24, 1998, hep-ph/9808414.

\bibitem{tournefier}
E.\ Tournefier (ALEPH Collab.), Report presented at the Int.\ Seminar
``Quarks '98'', Suzdal, May 17-24, 1998.

\bibitem{higgs}
F.\ Teubert, hep-ph/9811414;
G.\ D'Agostini and G.\ Degrassi, hep-ph/9902226.

\bibitem{vlf}
G.C.\ Branco and L.\ Lavoura, Nucl.\ Phys {B278} (1986) 738;
Phys.\ Lett.\ {B208} (1988) 123;
B.\ Mukhopadhyaya and S.\ Nandi, Phys.\ Rev.\ Lett.\  {66} (1991)
285; Phys.\ Rev.\ {D46} (1992) 5098;
W.S.~How, Phys.\ Rev.\ Lett.\ {69} (1992) 3587;
T.P.\ Cheng and L.F. Li, Phys.\ Rev.\ {D45} (1992) 1708;
G.C.\ Branco, T.~Morozumi, P.\ Parada and M.N.\ Rebelo, {\it ibid}
{D48} (1993) 1167; G.C.\ Branco,  P.\ Parada and M.N.\ Rebelo,
{\it ibid} {D52} (1995) 4217, hep-ph/9501347;
G.~Bhattacharyya, G.C.~Branco and W.S.~How, Phys.\ Rev.\ {D54} (1996)
2114,
hep-ph/9512239;
L.T.~Handoco and T.~Morozumi, Mod.\ Phys.\
Lett.\ {A10} (1995) 309, hep-ph/9409240; 
E.\ Ma, Phys.\ Rev.\ {D53} (1996) 2276, hep-ph/9510289;   
T.~Yoshikawa, Prog.\ Theor.\ Phys.\ {96} (1996) 269, hep-ph/9512251.
 
\bibitem{lavoura}
L.\ Lavoura and J.P.\ Silva, Phys.\ Rev.\ {D47} (1993) 1117.

\bibitem{n_families}
J.\ Shechter and J.W.F.\ Valle, Phys.\ Rev.\ {D21} (1980) 309;
{D22} (1980) 2227;
R.\ Mignani, Lett.\ Nuovo Cimento {28} (1980) 529; 
C.\ Jarlskog, Phys.\ Rev.\ Lett.\ {55} (1985) 1039; Z.\ Phys.\
{C29} (1985) 491; Phys.\ Rev.\ {D35} (1987) 1685;
M.\ Gronau, R.\ Johnson and  J.\ Shechter,  {\it ibid}  {D32}
(1985) 3062; 
M.\ Gronau, A.\ Kfir and R.\ Loewy, Phys.\ Rev.\ Lett.\ {56}
(1986) 1538; 
J.\ Bernabeu, G.C.\ Branco and M.\ Gronau, Phys.\ Lett.\ {B169}
(1986) 243; 
H.\ Harari and M.\ Leurer,  {\it ibid} {B181} (1986) 123; 
H.~ Frietzsch and J.\ Plankl, Phys.\ Rev.\ {D35} (1987) 1732;
J.F.\ Nieves and P.B.\ Bal, {\it ibid}  {D36} (1987) 315;
J.D.~Bjorken and I.~Dunitz, {\it ibid}  {D36} (1987) 2109.

\bibitem{santa}
A.\ Santamaria, Phys.\ Lett.\ {B305} (1993) 90, hep-ph/9302301.

\bibitem{fuji}
K.\ Fujikawa, Prog.\ Theor.\ Phys.\ {92} (1994) 1149;
hep-ph/9604358;
R.K.\ Babu, J.\ Pati and H.\ Stremnitzer, 
Phys.\ Rev.\ {D51} (1995) 2451, hep-ph/9409381;
H.\ Zheng, hep-ph/9602340.

\end{thebibliography}
\end{document}